\documentclass[conference,review]{IEEEtran}
\IEEEoverridecommandlockouts
\usepackage{cite}
\usepackage{multirow}
\usepackage{amsmath}
\usepackage{amssymb}
\usepackage{amsfonts}
\usepackage{algorithm}
\usepackage{algorithmic}
\usepackage{graphicx}
\usepackage{subfigure}
\usepackage{pifont}
\usepackage{caption}
\usepackage{textcomp}
\usepackage{xcolor}
\def\BibTeX{{\rm B\kern-.05em{\sc i\kern-.025em b}\kern-.08em
    T\kern-.1667em\lower.7ex\hbox{E}\kern-.125emX}}

\makeatletter
\newcommand{\linebreakand}{%
  \end{@IEEEauthorhalign}
  \hfill\mbox{}\par
  \mbox{}\hfill\begin{@IEEEauthorhalign}
}
\makeatother

\begin{document}

\title{Leveraging Latent Evolutionary Optimization for Targeted Molecule Generation\\
}

\author{
    \IEEEauthorblockN{Siddartha Reddy N\textsuperscript{*}}
    \IEEEauthorblockA{\textit{Applied Research, Quantiphi} \\
    siddartha.reddy@quantiphi.com}
    \and
    \IEEEauthorblockN{Sai Prakash MV\textsuperscript{*}}
    \IEEEauthorblockA{\textit{Applied Research, Quantiphi} \\
    mukkamala.prakash@quantiphi.com}
    \and
    \IEEEauthorblockN{Varun V}
    \IEEEauthorblockA{\textit{Applied Research, Quantiphi} \\
    varun.v@quantiphi.com}
    \and
    \IEEEauthorblockN{Saisubramaniam Gopalakrishnan\textsuperscript{†}}
    \IEEEauthorblockA{\textit{Applied Research, Quantiphi} \\
    gopalakrishnan.saisubramaniam@quantiphi.com}
    \and
    \IEEEauthorblockN{Vishal Vaddina}
    \IEEEauthorblockA{\textit{Applied Research, Quantiphi} \\
    vishal.vaddina@quantiphi.com}
    
    \thanks{\textsuperscript{*}Equal contribution}
    \thanks{\textsuperscript{†}Corresponding author}
    \thanks{Accepted as a paper at IEEE CEC 2024. Link to the full version will be provided with DOI once available.}
    
}

% \IEEEauthorblockA{\IEEEauthorrefmark{3} Affiliation 3}%

\maketitle

\begin{abstract}

Lead optimization is a pivotal task in the drug design phase within the drug discovery lifecycle. The primary objective is to refine the lead compound to meet specific molecular properties for progression to the subsequent phase of development.
In this work, we present an innovative approach, Latent Evolutionary Optimization for Molecule Generation (LEOMol), a generative modeling framework for the efficient generation of optimized molecules. LEOMol leverages Evolutionary Algorithms, such as Genetic Algorithm and Differential Evolution, to search the latent space of a Variational AutoEncoder (VAE). This search facilitates the identification of the target molecule distribution within the latent space. Our approach consistently demonstrates superior performance compared to previous state-of-the-art models across a range of constrained molecule generation tasks, outperforming existing models in all four sub-tasks related to property targeting. Additionally, we suggest the importance of including toxicity in the evaluation of generative models.
Furthermore, an ablation study underscores the improvements that our approach provides over gradient-based latent space optimization methods. This underscores the effectiveness and superiority of LEOMol in addressing the inherent challenges in constrained molecule generation while emphasizing its potential to propel advancements in drug discovery.
\end{abstract}


\section{Introduction}
In the initial stages of drug discovery, the primary objective is the synthesis and design of novel molecules with specific, desired properties. Currently, this task is both expensive and time-consuming, primarily due to the extensive size of the chemical space, estimated to range between $10^{23}$ and $10^{60}$ \cite{PMID:23963658} for drug-like molecules. One prevailing approach involves the high-throughput screening of large molecular libraries to identify hit molecules which exhibit desired properties. However, the computational intensity of screening large molecule libraries is a significant challenge.  An alternative to screening existing compounds is the de novo design of entirely novel molecules \cite{upm76087}. Early deep generative models \cite{pmlr-v80-jin18a, 10.1145/3394486.3403104,pmlr-v139-luo21a,NEURIPS2018_d60678e8,article,Shi*2020GraphAF:} have emerged as powerful tools for systematically exploring the vast chemical landscape. A conventional strategy involves constructing a generative model capable of translating a vector within a latent space into a molecule. Subsequently, exploration or optimization within this latent space 
helps uncover innovative molecules with desired characteristics. In this context, notations such as SMILES \cite{doi:10.1021/ci00057a005}, Molecular Fingerprints \cite{doi:10.1021/c160017a018}, Graphs \cite{4700287}, and SELFIES \cite{Krenn2019SelfreferencingES} have emerged to encode molecular structures into a machine-readable format. Models such as those proposed by Jin et al. \cite{pmlr-v80-jin18a} and Eckmann et al. \cite{pmlr-v162-eckmann22a} utilize graphs and SELFIES respectively for representing molecules. These models follow the aforementioned approach by encoding molecular representations into the latent space and conducting gradient descent optimization \cite{DeCao2018MolGANAI} searches within that space to generate desired molecules. 
Another conventional strategy is to utilize Reinforcement Learning for the generation of molecules with specific property constraints \cite{NEURIPS2018_d60678e8 , Shi*2020GraphAF:, article}. 
%Models such as  \cite{NEURIPS2018_d60678e8 , Shi*2020GraphAF:, article} employ Reinforcement Learning to optimize the properties of the generated molecular graph structures.
However, these models have not met the expected performance levels in generating molecules with desired properties.
% \textcolor{red}{as they have struggled to effectively incorporate non-differentiable oracles within the optimization process.}
% \textcolor{blue}{
%  The central goal in the early phases of Drug Discovery is to synthesize and design new molecules that exhibit specific, desired properties. At present this task is expensive in terms of cost and time due to the large size of the chemical space where the range of drug-like molecules has been estimated to be between $10^{23}$ and $10^{60}$. A common approach to do this task involves constructing a generative model that translates a position within a high-dimensional latent space into a molecule, followed by exploring or optimizing within this latent space to discover novel desired molecules. \cite{pmlr-v80-jin18a},\cite{pmlr-v162-eckmann22a} these models utilizes graphs and SELFIES to represent the molecule representation and subsequently follows the above approach by encoding the molecular representation into the latent space and perform gradient descent optimization search over the latent space to generate desired molecules. However, these models didn't meet the expected performance in generating molecules with desired properties because they couldn't leverage non-differentiable oracles within the optimization process.}

In this study, we present a pioneering methodology termed Latent Evolutionary Optimization for Molecule Generation (LEOMol). This innovative design is geared towards targeted molecule generation, wherein optimization over the latent space is directed by Evolutionary Algorithms \cite{7955308}. Evolutionary Algorithms are optimization techniques inspired by biological evolution, functioning through the maintenance of a population of candidate solutions. These solutions undergo iterative processes involving selection, recombination (crossover), and mutation operators, aiming to evolve towards improved solutions.
The LEOMol design intricately integrates the Variational AutoEncoder (VAE) \cite{Kingma2013AutoEncodingVB} framework with an optimization search process, leveraging Genetic Algorithm \cite{8862255} and Differential Evolution \cite{Storn1997DifferentialE}. The key characteristics of this work encompass:
 % \textcolor{blue}{Here we propose a novel approach Latent Evolutionary Optimization for Molecule Generation (LEOMol) a new design for targeted molecule generation whereas molecular optimization over the latent space is guided by the Evolutionary Algorithms. Evolutionary Algorithms (EAs) are optimization techniques inspired by biological evolution. They work by maintaining a population of candidate solutions and iteratively applying selection, recombination (crossover), and mutation operators to evolve towards better solutions. The LEOMol design integrates the Variational AutoEncoder (VAE) framework with an optimization search process that includes Genetic Algorithm (GA) and Differential Evolution (DE), both being subsets of Evolutionary Algorithms. Key features of LEOMol include:}
\begin{itemize}
    \item Its application of Genetic Algorithm (GA) and Differential Evolution (DE) search optimization in the latent space to produce drug-like molecules with desirable properties.
    \item Its capability to generate a substantial proportion of molecules tailored to specific properties while concurrently ensuring diversity and faster inference.
    \item Its proficiency in generating molecules possessing desired properties while preserving similarity to the input molecule — a pivotal aspect in Lead Optimization.
    % \item Its usefulness in drug discovery by facilitating the exclusive generation of non-toxic molecules. \textcolor{orange}{do we need this here?} or \textcolor{orange}{or replace with below point?}
    \item Introduce the significance of assessing the models with due consideration to the toxicity of the generated molecules.

    \item Its ability to leverage non-differentiable oracles such as RDKit within the search process in the latent space optimization for targeted molecule generation.
    %In this context, the SYN-FUSION model serves as the oracle function for predicting the toxicity of the candidate molecules.
\end{itemize}

% \textcolor{red}{I don't think we should mention SYN-FUSION here}
% \textcolor{red}{some citations missing}

% \textcolor{blue}{
%  \begin{itemize}
%     \item It's utilization of GA and DE search optimization on latent space to generate drug-like molecules with desirable properties.
%     \item It's ability to generate high percentage of property targeted molecules while ensuring both diversity of the molecules and swift inference.
%     \item It's ability to generate molecules with desired properties while maintaining similarity with the input molecule -- a crucial task in Lead Optimization
%     \item Addresses a critical requirement in drug discovery by enabling the exclusive generation of non-toxic molecules molecules. We here utilized the SYN-FUSION model as the oracle function for predicting the toxicity of the molecule candidates
% \end{itemize}
% }
The LEOMol model underwent evaluation across four distinct constrained molecule generation tasks pertinent to the Hit generation and Lead Optimization stages of Drug Discovery. In all tasks, LEOMol consistently demonstrated results that were either superior to or comparable with those achieved by state-of-the-art models.
\begin{figure*}[h]
    \centering
    \includegraphics[width=0.9\textwidth]{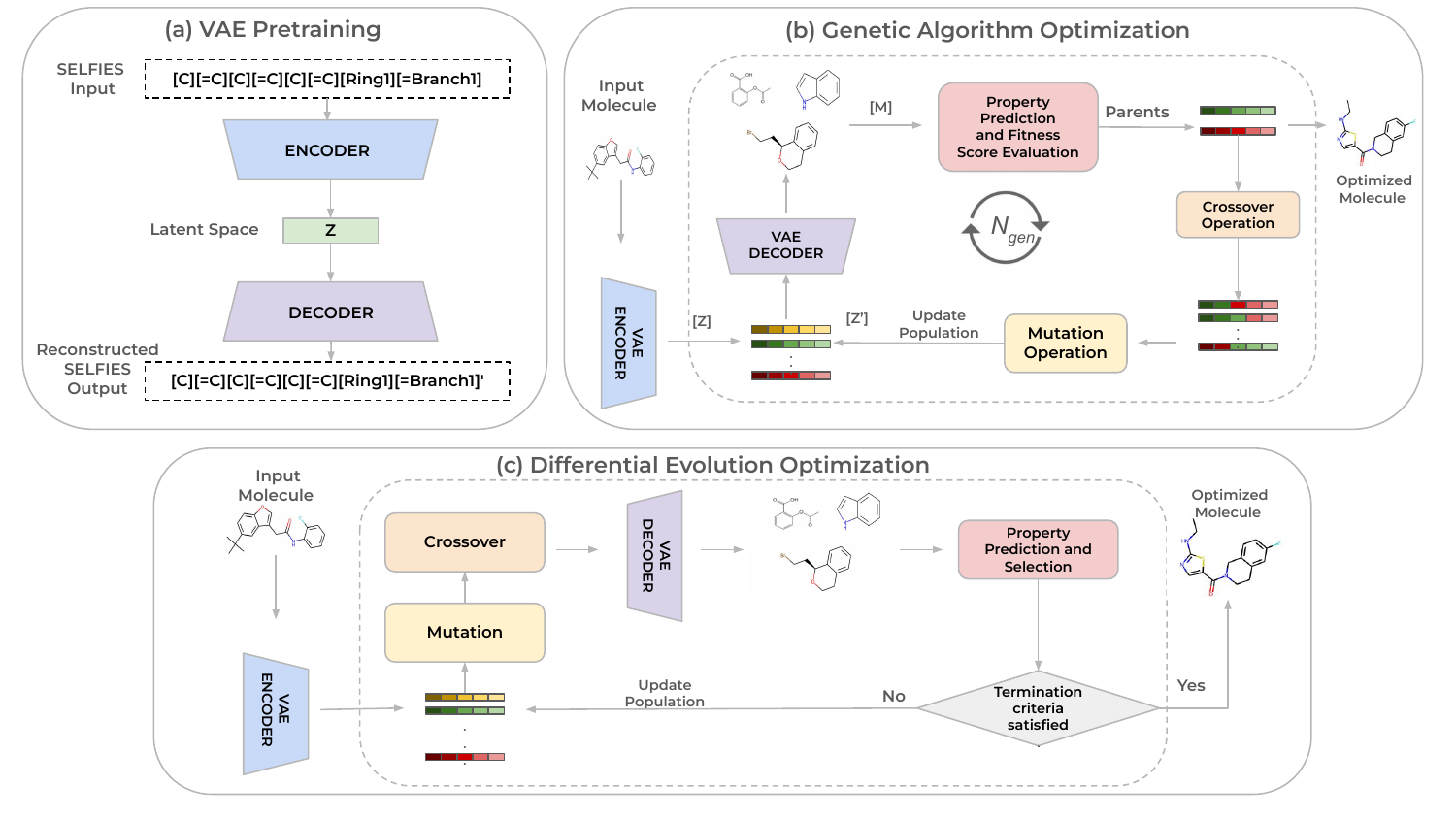}
    \caption{
    % Overview of the proposed LEOMol method: (a) We pre-train a Variational AutoEncoder (VAE) to learn to reconstruct and generate the drug-like molecule in SELFIES molecule representation. The pre-trained VAE is utilized with: (b) Genetic Algorithm and (c) Differential Evolution search strategies for exploring the latent space in pursuit of and optimization of the desired molecule.
Overview of the proposed LEOMol method:  (a) A Variational AutoEncoder (VAE) is pre-trained to acquire the ability to reconstruct and generate drug-like molecules using the SELFIES molecule representation. The pre-trained VAE is subsequently employed in conjunction with (b) Genetic Algorithm and (c) Differential Evolution search strategies to explore the latent space, aiming to optimize the desired molecule.}
    \label{fig:img1}
\end{figure*}

\section{Related Works}
% Related Works: A Survey of Approaches in Drug Discovery

% In the realm of drug discovery, the identification of potential drug targets marks the initiation of a multi-step process. One prevailing approach involves the high-throughput screening of compounds to identify hit compounds—molecules exhibiting affinity to the target protein. Experimental methods play a crucial role in this phase \cite{article1}, along with computational techniques such as docking \cite{SantosMartins2019AcceleratingAW, doi:10.1021/jm0306430} and molecular dynamics-guided binding free energy calculations \cite{doi:10.1021/acs.jcim.0c00116}.

Deep generative models leverage a learned latent space to represent the distribution of drug-like molecules. Early works such as Variational AutoEncoders (VAEs) \cite{doi:10.1021/acscentsci.7b00572}, encode SMILES strings to a continuous latent space. Addressing the limitations of SMILES-based representations, more recent models \cite{pmlr-v70-kusner17a, Dai2018SyntaxDirectedVA} incorporate rules into the decoding process, and \cite{pmlr-v162-eckmann22a} replace SMILES with SELFIES to ensure the generation of valid molecules. 
% To ensure chemical validity during generation,
Junction tree VAEs \cite{pmlr-v80-jin18a} use scaffold junction trees, offering a framework for the assembly of building blocks into valid molecular graphs. Although these approaches generate valid molecules, they encounter challenges in generating novel molecules with targeted property scores which are out of the training data distribution, due to the inability to incorporate non-differentiable oracles by design. 

Graph generative models, an alternative to sequence generation, have also been proposed.
% \cite{NEURIPS2018_1458e750, Simonovsky2018GraphVAETG,  Li2018MultiobjectiveDN, fu2022differentiable, 10.1145/3394486.3403104, pmlr-v139-luo21a, pmlr-v119-jin20a, NEURIPS2021_31445061}.
% \textcolor{violet}{\textbf{{related work section should have specific names of baselines. This look like extension of intro.}}}
These models often use surrogate models to predict molecular properties and guide optimization in the latent space. GCPN \cite{NEURIPS2018_d60678e8}, GraphDF \cite{pmlr-v139-luo21a}, MARS \cite{xie2021mars} and MolDQN \cite{article} employ Reinforcement Learning for the systematic creation or modification of molecular graphs. Typically, these methods construct molecules in a step-by-step manner, either atom by atom or fragment by fragment. Although effective in producing molecules with specific properties, these methods are computationally demanding, require numerous calls to property estimation functions, and also require task-specific finetuning.
%\cite{D1SC00231G,Nigam2019AugmentingGA}

Prior evolutionary algorithms function by generating the initial population of molecular representations, such as graph or text representations, emphasizing the execution of crossover and mutation operations directly on the input molecular representation itself. GB-GA \cite{C8SC05372C} designs molecules by performing crossover and mutation operations on the graph representations of the molecules. STONED \cite{D1SC00231G} 
works with the SELFIES representation for constrained molecule design. Despite its ability to generate molecules rapidly, this method still fails to meet the desired property constraints.

% Reinforcement learning-based methods directly optimize molecular properties through the systematic construction or alteration of molecular graphs. While powerful in generating molecules with desired properties, these methods are computationally intensive, requiring numerous calls to property estimation functions \cite{NEURIPS2018_d60678e8, article, xie2021mars, pmlr-v119-jin20b}. Despite challenges, these methods exhibit significant potential for generating molecules with tailored properties in drug discovery.

% \textcolor{violet}{\textbf{Compare and contrast with our proposed.}}

% Graph and RL generative models often generate molecules by atom and atom or fragment by fragment generation, whereas our approach generates the whole molecule at one shot generation and the existing VAE based models utilize gradient-based approaches for optimization search in latent space whereas we utilize Evolutionary Algorithm for optimization search in latent space. Also Existing Evolutionary Algorithms try to optimize the input molecule structure (Graph or Text) for constrained molecule design whereas our approach uses EAs to optimize the latent vector

% \textcolor{blue}{Current Graph and RL generative models for molecules typically generate molecules by atom and atom or fragment by fragment at a time. Similarly, most VAE-based models utilize gradient-based approach for optimizing the latent vector. Additionally, existing Evolutionary Algorithms (EAs) often focus on directly optimizing the molecule's structure, be it represented as a graph or text, to meet specific design constraints.}

In contrast, our approach takes a more holistic approach to molecule generation. Instead of building molecules atom by atom or fragment by fragment, we generate the entire molecule at once by decoding from the latent space. Furthermore, instead of relying on gradient-based methods, we harness the power of Evolutionary Algorithms (EAs) to optimize the latent vectors. This population-based approach leverages diverse candidate molecules and intelligently explores the landscape, leading to a wider range of potential solutions compared to gradient-based methods. Additionally, our Evolutionary Algorithms operate directly on the latent vectors, allowing for fine-grained control over the molecule's properties without directly manipulating its structure, avoiding the need for task-specific finetuning.

% \textcolor{orange}{I feel the below points are not required here, breaking the flow and unnecessary info} \textcolor{violet}{\textbf{{agreed}}} We employed neural network models to predict the toxicity of molecules. These models can take the form of graph-based networks \cite{Wang_2022} , transformer-based networks \cite{Irwin_2022} , or fusion \cite{10.1145/3584371.3613014} of both.

% This paper explores and compares the capabilities and limitations of both generative models and reinforcement learning-based methods in the context of drug discovery, shedding light on their applications and potential synergies in advancing the field.
% \subsection{Maintaining the Integrity of the Specifications}

% The IEEEtran class file is used to format your paper and style the text. All margins, 
% column widths, line spaces, and text fonts are prescribed; please do not 
% alter them. You may note peculiarities. For example, the head margin
% measures proportionately more than is customary. This measurement 
% and others are deliberate, using specifications that anticipate your paper 
% as one part of the entire proceedings, and not as an independent document. 
% Please do not revise any of the current designations.

\section{Methodology}
We introduce Latent Evolutionary Optimization for Molecule Generation (LEOMol), a novel methodology for constrained molecule design. This approach combines a Variational AutoEncoder (VAE) \cite{Kingma2013AutoEncodingVB} to learn the real-valued latent representation of drug-like molecules and employs Evolutionary Algorithms, primarily the Genetic Algorithm (GA) and Differential Evolution (DE), to search and create molecules within defined constraints. Figure \ref{fig:img1} illustrates the comprehensive workflow of our proposed method, LEOMol. 

% Here We introduced a new design for constrained molecule design by combining chemistry domain knowledge , A Variational AutoEncoder(VAE) to learn the real-valued latent representation of the drug like molecules and the Genetic Algorithm search for constrained molecule generation. Figure 1 illustrates the overall workflow of our approach LEOMol. We also utilize the RDKit tools for calculating the molecule related properites. 

% We opted for SELFIES as the molecular token representation over the SMILES molecular language due to its comprehensive robustness. SELFIES ensures that its every possible  string representation corresponds to a chemically valid molecule, thereby enabling the model to generate 100\% valid molecules. As the first step we pretrain the VAE with the ZINC250k dataset for comprehending both syntax and semantics of the SELFIES molecular language and then we combine the Genetic Algorithm search optimization with the pretrained VAE for desired molecule generation.
We choose SELFIES \cite{Krenn2019SelfreferencingES} as the molecular representation over SMILES due to its inherent robustness. 
% \textcolor{red}{SELFIES ensures that every possible string representation corresponds to a chemically valid molecule, thereby enabling the model to generate 100\% valid molecules.} Each point in the latent space of a VAE corresponds to a SELFIES representation which is always a valid molecule contrasting with SMILES where only fragmented, disconnected regions correspond to valid representations.

Each point in the latent space of a VAE corresponds to a SELFIES representation which is always a valid molecule. In contrast, there are only fragmented, disconnected regions corresponding to the valid representations in SMILES. In the initial step, we pre-train a VAE using the ZINC250k dataset to comprehend both the syntax and semantics of the SELFIES molecular language. Then, we integrate the Genetic Algorithm search optimization with the trained VAE to generate desired molecules. Additionally, we substitute the Genetic Algorithm with the Differential Evolution algorithm as an alternative search optimization strategy. We employ the open-source cheminformatics tool RDKit \cite{landrum2006rdkit} to compute molecule properties.
% \textcolor{orange}{can u explain the below para @sid?}Since there are non-differentiable tools like RDKit available and our approach doesn't necessarily require   we prefer using methods that don't rely on gradients. These methods can handle such tools well, allowing for a more accurate and efficient search compared to gradient-based approaches.

% \textcolor{orange}{Redundant}
% We utilized Genetic Algorithm (GA) and Differential Evolution(DE) as our optimization search algorithms due to their effectiveness in exploring and exploiting the latent space.

% \textcolor{violet}{\textbf{Since we are going to the methods from below, would suggest writing a line that in this work, we are considering both GA and DE for xyz reasons.}}

\subsection{\textbf{Variational AutoEncoder}}
A Variational AutoEncoder (VAE) functions as a generative model, acquiring a probabilistic representation of the training data. It employs an encoder neural network to map input data to a low-dimensional latent space and a decoder neural network to reconstruct the original data based on this latent space representation.
The initial step involves encoding a molecule $m$ into its SELFIES representation \{$x_1$, $x_2$, \ldots, $x_n$\}, where $n$ denotes the molecular length, and $x_i$ is selected from the predefined SELFIES vocabulary $V$ = \{$s_1$, $s_2$, \ldots, $s_d$\}, with $d$ representing the vocabulary size. A latent vector $z$ representing the encoded information of $m$ is passed to the VAE decoder to reconstruct it back, i.e., the conditional distribution $p(y | z)$, where $y$ = \{$x'_1$, $x'_2$, \ldots, $x'_n$\}, where $x'_i$ is the predicted SELFIES string. 
The pre-training follows the Evidence Lower Bound (ELBO) approach over molecules in the Zinc250k dataset that involves the calculation of:
1) Reconstruction Loss: The loss is determined by calculating the negative log-likelihood between the decoder outputs and the one-hot encoded representation of the input SELFIES string.
2) KL Divergence Loss: The loss is computed as a measure of the difference between the posterior distribution and a chosen prior distribution (typically Gaussian). The KL divergence loss encourages a more structured and interpretable latent space representation.
Upon the completion of the training process, the VAE exhibits the capability to generate novel drug-like molecules.
% within the input molecule $M$ undergoes embedding through the embedding layer, resulting in a vector representation $x$. The Variational AutoEncoder (VAE) is subsequently trained to encode $x$ into a latent space vector $z$ while simultaneously decoding it to $y$. The training process adheres to the Evidence Lower Bound (ELBO) approach for the VAE.

% % 1) **Reconstruction Loss:**
%    \[ \mathcal{L}_{rec} = -\mathbb{E}_{q_{\phi}(z|x)}[\log p_{\theta}(x|z)] \]

%    Here, \( q_{\phi}(z|x) \) is the posterior distribution, \( p_{\theta}(x|z) \) is the likelihood (decoder output), and the expectation is taken over the posterior distribution.

% % 2) **KL Divergence Loss:**
%    \[ \mathcal{L}_{KL} = \text{KL}(q_{\phi}(z|x) || p(z)) \]
%\textcolor{violet}{\textbf{read below and skip these two eqns.}}
   % This represents the Kullback-Leibler (KL) divergence between the posterior distribution \( q_{\phi}(z|x) \) and the chosen prior distribution \( p(z) \). The KL divergence encourages a more structured latent space.

% The overall ELBO loss is the sum of the reconstruction loss and the KL divergence loss:
% \[ \mathcal{L}_{ELBO} = - \mathcal{L}_{rec} + \mathcal{L}_{KL} \]
% \textcolor{violet}{\textbf{we have already added the negative sign for reconstruction loss, why do we need the negative sign again? Better to directly provide the ELBO loss and mention that the first part is reconstruction and second is the KL loss.}} 
% \textcolor{orange}{Missing eq number for some eqns}

\begin{equation}
 \mathcal{L}_{ELBO} = -\mathbb{E}_{q_{\phi}(z|x)}[\log p_{\theta}(x|z)] + \text{KL}(q_{\phi}(z|x) || p(z)) 
\end{equation}
where the first part is the loss for reconstruction, the second part is the KL loss, \( \phi \) represents the parameters of the encoder, \( \theta \) represents the parameters of the decoder, \( x \) is the input data, \( z \) is the latent variable, \( p(z) \) is the prior distribution, \( q_{\phi}(z|x) \) is the posterior distribution, \( p_{\theta}(x|z) \) is the likelihood (decoder output), and the expectation is taken over the posterior distribution, usually Gaussian over the latent space.

% These equations encapsulate the core principles of the ELBO loss used in training Variational AutoEncoders.
% Initially every string token in the input molecule $M$ will undergo the embedding layer to get a vector representation $x$ and we train the VAE to encode the $X$ to a latent space vector $z$ and simultaneously decode to $y$. We follow the Evidence Lower Bound(ELBO) training approach for our VAE. It consists of calculating two losses 1) Reconstruction loss is calculated by using negative log-likelihood between the decoder outputs and the one-hot encoded representation of the input SELFIES string 2) KL divergence loss is calculated as a measure of the difference between posterior distribution and chosen prior distribution(usually Gaussian) which promotes more structured and interpretable latent space representation. Once the training process is completed VAE is able to generate novel drug like molecules.

\subsection{\textbf{Evolutionary Algorithm Optimization}}

% Genetic Algorithms(GA) are optimization algorithms inspired by the process of natural selection and genetics. This algorithm reflects the process of natural selection where the fittest individuals are selected for reproduction in order to produce offspring of the next generation. 
Evolutionary Algorithms are optimization techniques inspired by natural selection and genetics. These algorithms emulate the process of natural selection, selecting individuals with the highest fitness for reproduction, thereby generating offspring in successive generations.

\begin{algorithm}

\caption{Genetic Algorithm for Molecule Optimization }
\label{alg:algorithm1}
\begin{algorithmic}[1]

\REQUIRE VAE encoder $f_{enc}$ , VAE decoder $f_{dec}$ , Fitness function $F(\cdot)$ , no.of generations $n_{gen}$ ,  crossover probability $p_c$ , mutation probability $p_m$ population size $n_m$ 

\STATE Initialize the population $Z$ \\
% \IF{GA}
\FOR{$i = 1$ to $n_{gen}$}
    \STATE $Parents$ $\leftarrow$ Choose fittest individuals from $Z$
    % \STATE $m$ = $f_{dec}(Z)$
    % \STATE Calculate the probability of each individual in $Z$
    % \STATE $P$ = $F(m)$
    % \STATE $Parents$ = Select($Z$ , $P$)
    \STATE $Z'$ = Crossover($Parents$ , $p_c$)
    \STATE $Z''$ = Mutation($Z'$ , $p_m$)
    % \STATE Update the new population
    \STATE $Z$ = $Z''$ $\leftarrow$ Update the population
\ENDFOR
 
\STATE $M$ $\leftarrow$ Choose the best individual from $Z$ 
% \STATE $m$ = $f_{dec}(Z)$
% \STATE $P$ = $F(m)$
% \STATE $M$ = SelectBest($Z$ , $P$)
% \label{Algorithm-1}

\end{algorithmic}
\end{algorithm}

\begin{algorithm}
\caption{Differential Evolution for Molecule Optimization}
\label{alg:algorithm2}
\begin{algorithmic}[1]

\REQUIRE VAE encoder $f_{enc}$ , VAE decoder $f_{dec}$ , Fitness function $F(\cdot)$ , no.of generations $n_{gen}$ ,  crossover probability $p_c$ , mutation probability $p_m$ , population size $n_m$ 
\STATE Initialize the population $Z$ \\

\FOR{i = 1 to $n_{gen}$}
    \STATE $Z'$ = \( \phi \)
    \FOR{$z$ in $Z$}
        \STATE a,b,c = random($Z$,3)
        \STATE $z'$ = Mutation($z$, a, b)
        \STATE $z''$ = Crossover($z'$, c)
        \STATE Insert($Z'$ , $z''$ )
    \ENDFOR
    \STATE $Z''$ = Select($Z$ , $Z'$) $\leftarrow$ Update the population

    \IF{best\_value($Z''$) satisfies constraint}
        \STATE break
    \ENDIF
   
\ENDFOR
\STATE $M$ $\leftarrow$ Choose the best individual from $Z$ 

\end{algorithmic}
\end{algorithm}
    %\STATE best\_value = findBestValue($Z''$)

% \STATE \textbf{Selection }Choose the best individual from the $Z$
% \STATE $M$ = $f_{dec}(Z)$
% \STATE $P$ = $F(M)$
% \STATE $M$ = SelectBest($Z$ , $P$)

% The pseudo-code in Algorithm \ref{Algorithm-1} outlines the functioning of the Genetic Algorithm (GA) in optimizing the latent space of a Variational AutoEncoder (VAE). Initially, the algorithm selects the initial population from encoded latent vectors or randomly sampled latent vectors $Z$, following a Normal distribution $N(0,1)$. The fittest latent vectors are determined by evaluating them using the fitness score function $F(\cdot)$. Subsequently, the algorithm employs crossover and mutation operations on these fittest individuals to generate a new population. This iterative process continues for $N_{gen}$ iterations.

\subsubsection{\textbf{Genetic Algorithm Approach}}

The pseudo-code specified in Algorithm \ref{alg:algorithm1} describes the operation of the Genetic Algorithm (GA) for optimizing a particular latent vector in the pre-trained Variational Autoendocer (VAE) which when decoded will generate a molecule that satisfies the desired properties. The algorithm selects an initial population from either, (i) randomly sampled latent vectors denoted as $Z = \{z_1, z_2, ..., z_n\}$, following a Normal distribution $N(0,1)$ or (ii) encoded latent vector $f_{enc}(m)$ if an input molecule $m$ is provided. The determination of the fittest latent vectors is achieved through their evaluation using the fitness score function $F(\cdot)$. Subsequently, the algorithm employs crossover and mutation operations in order over these fittest individuals, to generate a new population.

\textbf{Crossover} 
% \textcolor{violet}{\textbf{since this conference is evolutionary based, we can cutshort what is meant by crossover; two lines and then come to the orange part.; same for mutation}} 
% \textcolor{blue}{From a pool comprising the most suitable individuals, two parent individuals are chosen randomly. Subsequently, an index is randomly selected within the length of the latent space dimension. The segment corresponding to this index is then extracted from both parent individuals and merged to form a child individual. The execution of the crossover operation is contingent upon a specified crossover rate. If the randomly generated probability is below the crossover rate, the crossover is performed; otherwise, the child remains identical to the first parent.} This process is repeated until a new set of child individuals, matching the length of the original population, is generated.
% \begin{equation}
%   C = \begin{cases}
%   [P_1^{(1)}, P_1^{(2)}, ..., P_1^{(CP)}, P_2^{(CP+1)}, ..., P_2^{(L)}] & \text{if } r < CR \\
%   P_1 & \text{otherwise}
% \end{cases}
% \end{equation}
% Where $P_1^{(1)}$, $P_1^{(2)}$ are the parents randomly chosen from the elite list and $CR$ is the predefined crossover rate and $CP$ is the randomly selected crossover point within the gene length $L$ and $r$ is a random number uniformly distributed between 0 and 1.
Consider two parents, \(P_1\) and \(P_2\), selected randomly from an elite list. The crossover operation, denoted as \(C\), can be expressed as:
\begin{equation}
C_i = 
\begin{cases}
  P_{1i} & \text{if } r > CR \\
  P_{1i} & \text{if } r \leq CR \text{ and } i \leq CP \\
  P_{2i} & \text{if } r \leq CR \text{ and } i > CP
\end{cases}
\end{equation}

for \(i = 1, 2, \ldots, L\), where:
 \(C_i\) represents the \(i\)-th gene of the offspring \(C\), \(P_{1i}\) and \(P_{2i}\) are the \(i\)-th genes of parents \(P_1\) and \(P_2\), respectively, \(CR\) is the predefined crossover rate, \(CP\) is a randomly selected crossover point within the gene length \(L\), \(r\) is a random number uniformly distributed in the interval [0, 1].

This process is repeated until a new set of child individuals, matching the length of the original population, is generated.
% This formulation precisely outlines the crossover process, where the offspring gene \(C_i\) is determined based on the crossover rate \(CR\), the crossover point \(CP\), and the random variable \(r\). The expression ensures that the offspring inherits genetic material from both parents in a manner that is conditioned on the specified parameters.

\textbf{Mutation}
%From the pool of recently generated child individuals, mutations are applied probabilistically based on the mutation rate. if a randomly generated probability is less than the specified mutation rate, 
In the creation of the new population, a random gene is chosen within the genetic sequence length for each individual. Following this selection, a random value is drawn from a normal distribution with a standard deviation of 0.1 and added to the chosen gene, causing a minor modification in its value. This process is then replicated for each additional child individual in accordance with the specified mutation rate.

%This process introduces diverse genetic variations into the new population, facilitating exploration within the search space.

\begin{equation}
C^{(MP)} = \begin{cases}
  C^{(MP)} + \mathcal{N}(0, \sigma^2) & \text{if } r < MR \\
  C^{(MP)} & \text{otherwise}
\end{cases}
\end{equation}

Where $MR$ is the predefined mutation rate and $MP$ is the randomly selected mutation point within the gene length $L$
and $r$ is a random number uniformly distributed between 0 and 1.
% This iterative process persists for $N_{gen}$ iterations.
% From the pool of the fittest individuals, we randomly select two parent individuals. Then, we randomly pick an index within the latent space dimension's length. Next, we extract the segment referring to that index from both parent individuals and combine them to create a child individual. The crossover operation is performed based on a crossover rate. If the randomly generated probability is less than the crossover rate, perform crossover; otherwise, the child remains the same as the first parent. \textcolor{red}{need more brief on the slicing part} We repeat this process till we create a new set of child individuals with the length of the original population.

% \textbf{Mutation:}

% From the pool of newly created child individuals. We applied probabilistically based on the mutation-rate. For each individual in the population, if a randomly generated probability is less than the specified mutation rate, a random gene is selected within the genetic sequence length. Then, a small random value from a normal distribution with a standard deviation of 0.1 is added to the selected gene, inducing a slight change in its value. And we repeat this iteration for every other child individual. This process introduces diverse genetic variations into the new population, aiding exploration in the search space.

%This iterative process continues for $N_{gen}$ iterations.

\subsubsection{\textbf{Differential Evolution Approach}} The pseudo code specified in Algorithm \ref{alg:algorithm2} describes the operation of Differential Evolution (DE) algorithm. The DE algorithm, comparable to Genetic Algorithms (GA), operates within the domain of Evolutionary algorithms, entailing the iterative evolution of a solution population. However, DE distinguishes itself by its approach to updating the initial population, employing a sequential application of mutation and crossover strategies. The algorithm applies mutation and crossover operations to the initial population, subsequently replacing its individuals with the fittest candidates resulting from these operations.

\textbf{Mutation}
% \textcolor{violet}{\textbf{since this conference is evolutionary based, we can cutshort what is meant by crossover; two lines and then come to the orange part.; same for mutation}} 
% \textcolor{violet}{\textbf{why is it crossover first and then mutation in GA while it is the reverse in DE?}} 
% Mutation is executed iteratively for each individual within the population. For every individual, three random individuals (excluding the current individual) are selected. The mutation operation is then applied by adding a weighted difference between two of the three randomly chosen individuals earlier to the current individual, scaled by the mutation rate.
The iterative execution of mutation involves processing each individual within the population. For every individual, a selection is made of three random individuals (excluding the current one). Subsequently, the mutation operation is implemented by incorporating a weighted difference between two of the three randomly chosen individuals prior to the current individual, scaled by the mutation rate.

%This process introduces variation by altering the individual's genetic makeup based on the differences between randomly chosen individuals.

\begin{equation}
C' = C + MR \cdot (P_a - P_b)
\end{equation}

Where $C$ is the unique individual chosen from the population and %$MR$ is the predefined mutation rate 
and the $P_a$, $P_b$ are the two of the three randomly selected individuals.

\textbf{Crossover}
% Subsequently, the newly generated mutated individual is considered, and a set of indices across the gene length is chosen according to the crossover rate. The values at these indices are then swapped with those of a third randomly chosen individual, resulting in the final modified individual. This mutation and crossover operation are iteratively performed for each individual in the current population, collectively forming the new population. This iterative process continues for $n_{gen}$ iterations.
Following the generation of the mutated individual, a selection of indices spanning the gene length is made based on the crossover rate. The values at these selected indices are then interchanged with those of a third randomly chosen individual, culminating in the creation of the ultimately modified individual. This sequence of mutation and crossover operations is systematically applied to each individual within the existing population, collectively shaping the composition of the updated population. This iterative process persists for a total of \( n_{gen} \) iterations.

\begin{equation}
C'^{(k)} = \begin{cases}
  P_c^{(k)} & \text{if } r < CR \\
  C'^{(k)} & \text{otherwise}
\end{cases}
\end{equation}

% \textcolor{red}{We need to mention that this process iterates over k = 1 to k = L}

Where %$CR$ is the predefined crossover rate and 
$P_c$ is the third randomly selected individual and $r$ is a random number uniformly distributed between 0 and 1.

The iterative process involving mutation and crossover operations on latent vectors will persist until a superior latent vector aligning with the specified molecular properties is attained. The optimization procedure concludes upon the achievement of such a latent vector.
\begin{table*}[!t]
\normalsize
   \centering
    \caption{The table illustrates the top three molecules characterized by the highest Quantitative Estimate of Druglikeness (QED) and Penalized LogP scores produced by each model. The term LL denotes the application of a length limit constraint on the number of atoms within the molecules during the generation process. The baseline results are derived from \cite{pmlr-v162-eckmann22a}.}
    % \resizebox{0.8\textwidth}{!}{%
    % 
    % \resizebox{\textwidth}{!}{%
    % \setlength{\tabcolsep}{22pt}
    \begin{tabular}
    % [2\linewidth]
    {c|c|ccc|ccc}
    
    % \hline
    %     & \multicolumn{6}{c}{\textbf{Classification (Higher is Better)}} \\
    \hline
        % \multirow{2}{4em}
        \multirow{2}{4em}{ \centering Method} & 
        % \multirow{2}
        {\multirow{2}{4em}{\centering LL} } & \multicolumn{3}{c|}{Penalized logP} & \multicolumn{3}{c}{QED}  \\ 
        & & 1st & 2nd & 3rd & 1st & 2nd & 3rd \\
       \hline 
        ZINC & - & 4.52 & 4.30 & 4.23 & 0.948 & 0.948 & 0.948  \\ 
        JT-VAE & \ding{55} & 5.3 & 4.93 & 4.49 & 0.925 & 0.911 & 0.91  \\ 
        GRAPHDF & \ding{55} & 13.7 & 13.2 & 13.2 & \textbf{0.948} & \textbf{0.948} & \textbf{0.948}  \\ 
        MARS & \ding{55} & \textbf{45} & \textbf{44.3} & \textbf{43.8} & \textbf{0.948} & \textbf{0.948} & \textbf{0.948}  \\ 
        \hline
        GCPN & \checkmark & 7.98 & 7.85 & 7.8 & \textbf{0.948} & 0.947 & 0.946  \\
        MOLDQN & \checkmark & 11.8 & 11.8 & 11.8 & \textbf{0.948} & 0.943 & 0.943  \\ 
        LIMO & \checkmark & 10.5 & 9.69 & 9.6 & 0.947 & 0.946 & 0.945  \\ 
        \hline
        \textbf{LEOMol\textsubscript{GA}} & \checkmark & \textbf{18.48} & \textbf{18.12} & \textbf{17.8} & \textbf{0.948} & \textbf{0.948} & 0.946  \\ 
        \textbf{LEOMol\textsubscript{DE}} & \checkmark & 16.32 & 16.11 & 16.04 & 0.945 & 0.945 & 0.942  \\ 
        \hline

    \end{tabular}
    % }
    \label{table:1}
\end{table*}

\begin{table*}
\normalsize
 %   \captionsetup{font=large}
    \caption{
A comparative analysis of models to evaluate their diversity, quantified by assessing the deviation from the average pairwise Tanimoto similarity between Morgan fingerprints. The Success(\%), represents the proportion of generated molecules situated within the predetermined target range. This examination encompassed four distinct property targeting tasks.}
    \centering
    % \resizebox{\textwidth}{!}{%
    \begin{tabular}[width=2\linewidth]{c|cc|cc|cc|cc}

%      \hline
%         \multirow{2}{4em}{\centering Method}  & \multicolumn{2}{c}{-2.5 \leq logP \leq -2} & \multicolumn{2}{c}{5 \leq logP \leq 5.5} & \multicolumn{2}{c}{150 \leq MW \leq 200} & \multicolumn{2}{c}{500 \leq MW \leq 550} \\ 
%         & Success (\%) & Diversity & Success (\%) & Diversity & Success (\%) & Diversity & Success (\%) & Diversity\\
%        \hline 
%         ZINC & 0.3 & 0.919 & 1.3 & 0.909 & 1.7 & 0.938 & 0 & - \\ 
%         JT-VAE & 11.3 & 0.846 & 7.6 & 0.907 & 0.7 & 0.824 & 16 & 0.898 \\ 
%         ORGAN & 0 & - & 0.2 & \textbf{0.909} & 15.1 & 0.759 & 0.1 & \textbf{0.907}  \\
%         GCPN & 85.5 & 0.392 & 54.7 & 0.855 & 76.1 & 0.921 & \textbf{74.1} & 0.92  \\ 
%         LIMO & 10.4 & \textbf{0.914} & 5 & 0.923 & 7 & 0.907 & 2 & 0.908  \\ 
%         \hline 
%         \textbf{LEOMol\textsubscript{GA}} & \textbf{89} & 0.911 & \textbf{99.7} & 0.884 & \textbf{99.8} & 0.909 & \textbf{74.2} & 0.881 \\ 
%         \textbf{LEOMol\textsubscript{DE}} & \textbf{99} & 0.910 & \textbf{100} & 0.892 & \textbf{100} & 0.907 & \textbf{88} & 0.883 \\ 
%         \hline
%     \end{tabular}
% % }
%     \label{table:2}
% \end{table*}

 \hline
        \multirow{2}{*}{\centering Method} & \multicolumn{2}{c|}{$-2.5 \leq \log P \leq -2$} & \multicolumn{2}{c|}{$5 \leq \log P \leq 5.5$} & \multicolumn{2}{c|}{$150 \leq \text{MW} \leq 200$} & \multicolumn{2}{c|}{$500 \leq \text{MW} \leq 550$} \\ 
        & Success (\%) & Diversity & Success (\%) & Diversity & Success (\%) & Diversity & Success (\%) & Diversity\\
        \hline 
        ZINC & 0.3 & 0.919 & 1.3 & 0.909 & 1.7 & 0.938 & 0 & - \\ 
        JT-VAE & 11.3 & 0.846 & 7.6 & 0.907 & 0.7 & 0.824 & 16 & 0.898 \\ 
        ORGAN & 0 & - & 0.2 & \textbf{0.909} & 15.1 & 0.759 & 0.1 & \textbf{0.907}  \\
        GCPN & 85.5 & 0.392 & 54.7 & 0.855 & 76.1 & 0.921 & \textbf{74.1} & 0.92  \\ 
        LIMO & 10.4 & \textbf{0.914} & 5 & 0.923 & 7 & 0.907 & 2 & 0.908  \\ 
        \hline 
        \textbf{LEOMol\textsubscript{GA}} & \textbf{89} & 0.911 & \textbf{99.7} & 0.884 & \textbf{99.8} & 0.909 & \textbf{74.2} & 0.881 \\ 
        \textbf{LEOMol\textsubscript{DE}} & \textbf{99} & 0.910 & \textbf{100} & 0.892 & \textbf{100} & 0.907 & \textbf{88} & 0.883 \\ 
        \hline
    \end{tabular}
    \label{table:2}
\end{table*}

\section{Experiments}

% To demonstrate the effectiveness of our approach LEOMol on  effective  molecular generation we evaluated our model on the following constrained molecule generation objectives in comparison with the state-of-the-art generative models.
To demonstrate the efficacy of our LEOMol approach in the realm of molecular generation, we evaluated our model against the state-of-the-art generative models across various molecular generation objectives:

\textbf{a) Property Optimization} 
The primary aim here is to generate molecules while maximising property scores. This holds particular significance in the field of drug development, where the objective is to identify molecules characterized by highly optimized properties of interest.

\textbf{b) Property Targeting} 
The objective here is to generate molecules satisfying property scores within a predefined interval. This is essential in the creation of virtual libraries of molecules designed to align with specific molecular properties.

\textbf{c) Constrained Property Optimization} 
The objective here is to generate novel molecules that simultaneously adhere to a predefined similarity threshold with the input molecules, while also enhancing the desired properties. This becomes particularly crucial during lead optimization, where the objective is to refine the molecular properties of lead molecules while preserving their structural similarity.
% The objective here is to generate novel molecules with high property scores. This is crucial in drug development where the application is to find molecules with highly optimized properties of interest.

% The objective here is to generate novel molecules with property scores within a specified range. This type of ability is crucial when generating virtual libraries of molecules tailored to a particular property of molecule.

% The objective here is to generated novel molecules while maintaining a specific similarity threshold with the input molecules and improving the desired properties. This is important when we have to do the lead optimization where we need to improve the molecular property of the lead molecules while maintaining the structural similarity. 

% The objective here is to generate novel Non-Toxic molecules. Generating molecules with less toxicity accelerates the drug development process by enabling quicker progression to human trials.

\textbf{Dataset} 
The Variational AutoEncoder (VAE) was pretrained with the ZINC250k molecule dataset. The optimization tasks discussed in this section utilized the same dataset.
% We have pretrained the VAE with the ZINC250k molecule dataset which contains 250,000 drug molecules. We also utilized the same dataset for all the optimization tasks mentioned above.

\textbf{Configuration}

In line with the experimental setup of the LIMO \cite{pmlr-v162-eckmann22a} model, the Variational AutoEncoder (VAE) was pre-trained for 18 epochs using the Adam Optimizer with a learning rate of 0.001. 
% The Variational AutoEncoder (VAE) underwent 32 epochs of training, utilizing a learning rate of 0.001 and the Adam Optimizer.
The experiments consistently utilized the same pre-trained VAE and maintained a latent space dimension of 1024. 
% The Genetic Algorithm employed a crossover rate of 0.8 and a mutation rate of 0.4, while Differential Evolution utilized a crossover rate of 0.5 and a mutation rate of 0.8. 
After performing hyperparameter optimization, we determined a crossover rate of 0.8, a mutation rate of 0.4, a total of 20 generations and a population size of 20 for optimizing penalized logP. Subsequently, these hyperparameters were applied uniformly across all additional property optimization tasks.
% We performed hyperparameter tuning to choose these parameters. 
We customized uniquely tailored fitness scoring functions for each experiment, aligning precisely with the specific task requirements outlined in the provided problem statement. For predicting molecule properties, we utilized RDKit as our property predictor. All experiments were conducted on a single T4 GPU and 8 CPU cores with 52 GB of memory.
% The VAE underwent 32 epochs of training, utilizing a learning rate set at 0.001 and employing the Adam Optimizer, while following the ELBO optimization strategy. Consistently, the experiments employed the identical pretrained VAE weights and maintained a latent space dimension of 1024. Both GA and DE approaches have been implemented for each task. 

% Moreover, the Genetic Algorithm (GA) employed a crossover rate of 0.8 and a mutation rate of 0.4. And Differential Evolution (DE) utilized a crossover rate of 0.5 and a mutation rate of 0.8, both applied over 20 generations with a population size of 100. 
% Customized fitness scoring functions were tailored uniquely for each experiment, aligning precisely with the specific task requirements outlined in the given problem statement. 

\renewcommand{\arraystretch}{1.2}
%\vspace{2cm}
\begin{table*}[h]
\normalsize
 %   \captionsetup{font=large}
    \caption{Performance in similarity-constrained penalized logP maximization, for each method and minimum similarity threshold ($\delta$), the average improvement and standard deviation are provided for molecules meeting the similarity threshold. Additionally, the percentage of refined molecules adhering to the similarity constraint (\%SUCC.) is included. The baseline results are sourced from \cite{pmlr-v162-eckmann22a}.} 
    \centering
    \resizebox{\textwidth}{!}{%
    \begin{tabular}{c|cc|cc|cc|cc|cc|cc}
    
    \hline
        \multirow{2}{*}{$\delta$}  & \multicolumn{2}{c|}{JT-VAE} & \multicolumn{2}{c|}{GCPN} & \multicolumn{2}{c|}{MOLDQN} & \multicolumn{2}{c|}{LIMO} & \multicolumn{2}{c|}{\textbf{LEOMol\textsubscript{GA}}} & \multicolumn{2}{c}{\textbf{LEOMol\textsubscript{DE}}}\\ 
        & IMPROV. & \%SUCC & IMPROV. & \%SUCC & IMPROV. & \%SUCC & IMPROV. & \%SUCC & IMPROV. & \%SUCC & IMPROV. & \%SUCC\\
       \hline 
        0.0 & 1.9 ± 2.0 & 97.5 & 4.2 ± 1.3 & \textbf{100} & 7.0 ± 1.4 & \textbf{100} & 10.1 ± 2.3 & \textbf{100} & \textbf{13.73 ± 1.72} & \textbf{100} & 12.47 ± 1.14 & \textbf{100}\\ 
        0.2 & 1.7 ± 1.9 & 97.1 & 4.1 ± 1.2 & \textbf{100} & 5.1 ± 1.8 & \textbf{100} & 5.8 ± 2.6 & 99 & \textbf{8.68 ± 1.7} & 99 & \textbf{9.31 ± 1.63} & 97 \\ 
        0.4 & 0.8 ± 1.5 & 83.6 & 2.5 ± 1.3 & \textbf{100} & 3.4 ± 1.6 & \textbf{100} & 3.6 ± 2.3 & 93.7 & \textbf{5.8 ± 1.93} & 90  &
     \textbf{5.95 ± 1.8} & 81  \\
    
    \hline
      
    \end{tabular}
}
    \label{table:3}
\end{table*}

\begin{table*}
\normalsize
%    \captionsetup{font=large}
    \caption{Percentage of generated toxic molecules during the optimization process for QED, both with and without the inclusion of a toxicity constraint.}
    \centering
    \resizebox{1\textwidth}{!}{%
    \begin{tabular}[width=2\linewidth]{c|ccc|ccc}
    
    \hline
        \multirow{2}{10em}{\centering Task}  & \multicolumn{3}{c}{LEOMol\textsubscript{GA}} & \multicolumn{3}{c}{LEOMol\textsubscript{DE}}\\        
        
     & Non-Toxic Molecules(\%) & Diversity & QED Score & Non-Toxic Molecules(\%) & Diversity & QED Score\\
       \hline 
        % Toxicity Optimization & 98 & 0.898 & 0.4264 ± 0.1708 & 100 & 0.887 & 0.4341 ± 0.1712\\ 
        w/o. Constraint & 18 & 0.882 & 0.8602 ± 0.0286 & 31 & 0.892 & 0.822 ± 0.0317 \\
        w. Constraint & 100 & 0.883 & 0.8204 ± 0.0352 & 100 & 0.885 & 0.7782 ± 0.0557\\ 
        
    \hline
      
    \end{tabular}
}
    \label{table:4}
\end{table*}
\textbf{Baselines} A comparative analysis was conducted between the implemented methodology and several state-of-the-art molecular generation algorithms, namely JT-VAE \cite{pmlr-v80-jin18a} and LIMO \cite{pmlr-v162-eckmann22a}. These algorithms utilize Variational AutoEncoders (VAEs) for the generation of molecular graphs and SELFIES representations, respectively. Furthermore, a comparison was undertaken with GCPN \cite{NEURIPS2018_d60678e8}, GraphDF \cite{pmlr-v139-luo21a}, and MolDQN \cite{article}. These approaches employ reinforcement learning strategies, with the reward computed using RDKit, for the generation of target molecules.

% We performed a comparative analysis between our methodology and various state-of-the-art molecular generation algorithms, including JT-VAE \cite{pmlr-v80-jin18a} and LIMO \cite{pmlr-v162-eckmann22a}, which utilize Variational AutoEncoders (VAEs) for generating molecular graphs and SELFIES representations, respectively. Additionally, we compared our approach with GCPN \cite{NEURIPS2018_d60678e8}, GraphDF \cite{pmlr-v139-luo21a}, and MolDQN \cite{article}, all of which employ reinforcement learning strategies with the reward computed using RDKit for the generation of target molecules.
% We conducted a comparative analysis between our approach and several state of the art molecular generation algorithms such as \textcolor{orange}{the portion before the colon  and the follow up after the colon don't sync well - maybe we can add several SOTA mol gen algorithms such as ...} \textcolor{blue}{OK} JT-VAE \cite{pmlr-v80-jin18a}, LIMO \cite{pmlr-v162-eckmann22a} utilizing VAEs for generating molecular graphs and SELFIES representation respectively—alongside GCPN \cite{NEURIPS2018_d60678e8}, GraphDF \cite{pmlr-v139-luo21a}, MolDQN \cite{article} all employing reinforcement learning strategies for the generation of target molecules.

% \subsection{Results}
\subsection{Property Optimization} 
% The primary aim is to generate molecules exhibiting elevated property scores. This holds particular significance in the field of drug development, where the objective is to identify molecules characterized by highly optimized properties of interest.
\textbf{Setup}
To demonstrate the effectiveness of LEOMol in optimizing properties, the focus was directed towards the maximization of two molecular attributes: Penalized LogP (P-LogP) \cite{pmlr-v80-jin18a}  and Quantitative Estimate of Drug-likeness (QED) \cite{a15f1730fe1a4335b2bca04d28d66d36}. LogP represents the logarithm of the partition coefficient of a solute between octanol and water. P-LogP is derived by subtracting the Synthetic Accessibility (SA) score and the count of long cycles from the LogP score. The QED score serves as an indicator of drug-likeness. The P-LogP metric exhibits an unbounded range, whereas the QED score ranges between 0 and 1, with the maximum known QED score being 0.948. The determination of the fitness score for each molecule was based on its property score.
% To exhibit the efficacy of LEOMol on property optimization, we focused on maximizing two molecular properties: Penalized LogP (P-LogP)\cite{pmlr-v80-jin18a} and QED \cite{a15f1730fe1a4335b2bca04d28d66d36}. LogP represents the logarithm of the partition ratio of a solute between octanol and water, whereas P-LogP is derived by subtracting the Synthetic Accessibility (SA) score and the count of long cycles from the LogP score. QED score serves as an indicator of drug-likeness. It is noteworthy that the computation of both properties was carried out using RDKit.
% The P-LogP metric exhibits an unbounded range, while the QED score ranges between 0 and 1, with the maximum known QED score being 0.948 \cite{fang2023domainagnostic}. The fitness score for each molecule was determined based on its property score.

% For this task, we employed a pretrained Variational AutoEncoder (VAE) and implemented both GA and DE strategies, each consisting of 20 iterations to generate 200 molecules.

\begin{equation}
\textbf{fitness score} = P_{s}(m)
\end{equation}

where $m$ is the generated molecule and $P_{s}(\cdot)$ is the property predictor.

% To establish a benchmark, we took reference baseline results from \cite{pmlr-v162-eckmann22a} \textcolor{orange}{reference needed?} \textcolor{blue}{yes}. This approach ensures a comprehensive evaluation of our method's \textcolor{orange}{method's performance - doesn't go well with each other} \textcolor{blue}{I feel this approach ensures statement is not needed we can skip it} performance relative to established practices in the field.
% In this task we choose Penalized LogP and QED molecular properties to be maximized. LogP is the logarithm of the partition ratio of the solute between octanol and water wheras Penalized LogP is the LogP score minus the synthetic accessibility (SA) score and the number of long cycles. And QED score is the indicator of drug-likeness. Note that we used the RDKit to compute both the properties. Penalized LogP metric has an unbounded range while QED has a range of [0,1] and the maximum QED score known in 0.948. To perform this task we used the pretrained VAE and performed Genetic Algorithm 20 iterations and generated 200 molecules with the fitness score being the property score of the molecule. We report the top 3 scores over the generated molecules and time taken by each method. We took the baseline results from the LIMO paper.

\textbf{Results}
% \textcolor{red}{In presenting the results, the top three scores among the generated molecules are provided. }
Table \ref{table:1} presents the outcomes related to the top three unique and high-property molecules identified by each model. Notably, the P-LogP metric shows a significant dependence on the molecule's length. Unlike models such as GCPN, MOLDQN, and LIMO, which adhere to length constraints, the LEOMol approach demonstrates notable advancements, particularly in the task of P-LogP maximization. It is important to highlight that the MARS model, characterized by the absence of length restrictions, exhibits the capability to generate extended carbon chains, thereby yielding heightened P-LogP scores.
% In presenting the outcomes, we provide the top three scores among the generated molecules. Table \ref{table:1} presents the outcomes pertaining to the top three unique and high-property molecules identified by each model. It is noteworthy that the P-LogP metric exhibits a substantial dependency on the molecule's length. In contrast to models such as GCPN, MOLDQN, and LIMO, which adhere to length constraints, our approach LEOMol demonstrates noteworthy advancements, particularly in the task of P-LogP maximization. It is noteworthy that the MARS model, characterized by the absence of length restrictions, exhibits the capability to generate extended carbon chains, thereby yielding heightened P-LogP scores.
% Table \ref{table:1} shows the results of the top 3 unique and high property molecules found by each model. Note that the P-logP is almost entirely dependent on the molecule length. In comparison with the models (GCPN, MOLDQN, LIMO) which follow length limit our approach shows significant improvement over those models especially for Penalized logP maximization task. MARS, without a length restriction, can produce extended carbon chains, leading to elevated P-LogP scores.

\subsection{Property Targeting} 
\textbf{Setup} In the task, two molecular properties, specifically LogP and Molecular Weight, were identified. Distinct property ranges were subsequently assigned for each property. The prescribed target ranges are articulated as follows: $-2.5 \leq \text{LogP} \leq -2$, $5 \leq \text{LogP} \leq 5.5$, $150 \leq \text{MW} \leq 200$, and $500 \leq \text{MW} \leq 550$. Evaluation of the task is based on two primary metrics: Success (\%) and Diversity.
Success (\%) is determined by calculating the percentage of generated molecules that adhere to the specified property target constraints out of the total number of generated molecules. Diversity, conversely, is computed as one minus the average pairwise Tanimoto similarity between the Morgan fingerprints \cite{doi:10.1021/ci100050t} of the generated molecules. To perform this assessment, a set of 100 molecules was generated, and subsequently, both Success (\%) and Diversity values were computed based on the defined property constraints.
This thorough analysis yields insights into the effectiveness and diversity of the generated molecules concerning the specified LogP and Molecular Weight ranges.
Within the fitness function, an additional variable denoted as $t$ was incorporated into the scoring function, representing the midpoint of the property range. 
\begin{equation}
\textbf{fitness score} = P_{s}(m) - t
\end{equation}

% In the fitness function, We introduced an additional variable denoted as $y$ into the scoring function, representing the midpoint of the property range. 
\textbf{Results}
The results of the LEOMol model, in comparison with state-of-the-art models pertaining to this specific task, are delineated in Table \ref{table:2}. The outcomes indicate that the proposed approach surpasses other models significantly across all four sub-tasks, manifesting commendable success rates and comparable diversity scores. Particularly noteworthy is the performance in sub-tasks characterized by $5 \leq \text{logP} \leq 5.5$ and $150 \leq \text{MW} \leq 200$, where our method exhibited a notable 82.8\% and 31.4\% relative improvement in Success (\%) compared to the second-best performing model (GCPN). Furthermore, our approach showcased substantial advancements over other state-of-the-art Variational AutoEncoder (VAE) models, namely JT-VAE and LIMO.
% The outcomes of the LEOMol model, in comparison with state-of-the-art models for this particular task, are detailed in Table \ref{table:2}. It is evident from the results that our approach significantly outperformed other models in all four sub-tasks, exhibiting high success rates and comparable diversity scores. Notably, in the sub-tasks characterized by $5 \leq \text{logP} \leq 5.5$ and $150 \leq \text{MW} \leq 200$, our method demonstrated a remarkable 82.8\% and 31.4\% relative improvement on Success (\%) over the second-best performing model (GCPN). Moreover, our approach demonstrated substantial advancements over other state-of-the-art Variational AutoEncoder (VAE) models, specifically JT-VAE and LIMO.
% We incorporated an additional variable, $y$, within the score function, which represents the midpoint of the property range.

% Table \ref{table:2} shows the results of the LEOMol model with comparison with the state of the art models for this task. We can clearly observe our approach performed significantly over the other models in 3 out of 4 sub-tasks with high success(\%) and comparable diversity scores. Especially coming to the $5 \leq logP \leq 5.5$, $150 \leq MW \leq 200$ sub-tasks our method has a 82.8\% and 35\% relative improvement with the second best performing model(GCPN). And our method has significant improvement over the other state-of-the-art VAE models(JT-VAE, LIMO).

\subsection{Constrained Property Optimization} 
\textbf{Setup} In this experimental setting, a total of 800 molecules were systematically chosen based on their lowest P-LogP scores from the ZINC250k dataset, following the assessment methodology elucidated in \cite{pmlr-v80-jin18a}. The primary aim at this stage is to generate innovative molecules characterized by heightened P-LogP scores, all while maintaining a significant degree of resemblance to the aforementioned set of 800 molecules. The quantification of similarity is accomplished through the application of the Tanimoto similarity metric, which involves a comparative analysis of the Morgan fingerprints associated with each molecule.
To effectively address this dual objective, adjustments were made to the computation of the fitness score. This adaptation involved the integration of a similarity constraint, denoted as $\text{sim}(G, G_0)$, into the fitness function.

\begin{equation}
\textbf{fitness score} = 
\begin{cases}
    P_{s}(M), & \text{if }  \text{sim}(G, G_0) \geq \delta \\
    P_{s}(M) - \lambda * (\delta - \text{sim}) & \text{if }  \text{sim}(G, G_0) < \delta 
\end{cases}
\end{equation}

% The mean and standard deviation of the highest P-LogP improvement scores, along with the success rate (expressed as a percentage, indicating the number of molecules out of 800 that were able to improve while adhering to the similarity threshold),\textcolor{orange}{isn't it better to mention this info in the table rather than here?} \textcolor{blue}{We mentioned the task objective but we didn't mentioned about the 800 mols, I thought if I mentioned it there it will become too big for a caption} were calculated over the 800 molecules. 
\textbf{Results}
The results are systematically presented in Table \ref{table:3}. Despite not achieving a 100 percent success rate within the (0.2 and 0.4) similarity constraints, our approaches consistently demonstrated substantial improvements in P-LogP across all similarity thresholds. This highlights the reliable performance of LEOMol, even in environments characterized by multiple constraints.
% The results are systematically presented in Table \ref{table:3}. Although our approaches did not achieve a 100 percent success rate within the (0.2 and 0.4) similarity constraints, it has consistently shown the most substantial P-LogP improvements across all similarity thresholds. This underscores the consistent performance of the LEOMol, even in environments characterized by multiple constraints.

% Over the 800 molecules the mean and standard deviation of the highest p-logp improvement scores and the success(\%)(Out of 800 molecules how many molecules were able to improve while maintaining similarity threshold) are calculated and reported in the Table \ref{table:3} while our approach may not able to maintain the 100\% success rate in the (0.2,0.4) similarity constraints but it has achieved highest p-logp improvements across all similarity thresholds and a significant improvement over the other state-of-the-art models. This shows the ability of the LEOMol to perform consistently even in the multiple constraints environment.

\section{Analysis}
 \subsection{Toxicity Optimization}
 % In this context, we introduce a novel optimization task focused on enhancing molecules while ensuring their non-toxicity. Incorporating toxicity considerations into the optimization process is imperative, as molecules, despite meeting all property constraints, are deemed unsuitable for progression into subsequent stages of drug discovery if they exhibit toxicity.
 
In this context, a novel optimization task is introduced, emphasizing the improvement of molecular properties while concurrently ensuring their non-toxic nature. The inclusion of toxicity considerations in the optimization process is essential since molecules, even when compliant with all property constraints, are deemed unsuitable for advancement to subsequent stages of drug discovery if they demonstrate toxicity. So, we endeavor to optimize molecules while ensuring the containment of their toxicity through the utilization of LEOMol.
% To systematically evaluate the capabilities of the LEOMol model in both single and multi-objective optimization tasks, we divided the task into three distinct sub-tasks:
% 1) Non-Toxic Molecule Generation: This sub-task focuses solely on generating non-toxic molecules as the primary objective.
% 2) QED Optimization: Here, the exclusive objective is to generate molecules with higher QED scores.
% 3) Non-Toxic QED Optimization: The objective in this sub-task is to concurrently generate non-toxic molecules with higher QED scores.
The significance of incorporating a toxicity constraint in property optimization was examined through an experiment. This experiment involved the generation of molecules with elevated QED scores both with and without the imposition of a toxicity constraint. The assessment of these tasks relies primarily on three metrics: the proportion of non-toxic molecules, the diversity observed among the generated molecules, and the standard deviation and mean values of QED scores for the generated molecules. A collection of 100 molecules was generated for each sub-task, and the fitness score varied correspondingly based on the specific objective in question.
% The evaluation of these tasks is predominantly based on three metrics: the percentage of non-toxic molecules, the diversity of generated molecules, and the standard deviation and mean values of QED scores for the generated molecules. For each sub-task, we generated a set of 100 molecules, and the fitness score varied accordingly for each specific objective.
To predict the toxicity of the generated molecules, we employed a neural network trained on the MoleculeNet dataset, which forms an integral component of the fitness score function. Due to the novelty of this task, there are presently no established baselines for conducting comparative analyses.
% In this task the aim is to generate non-toxic molecules. We utilized SYN-FUSION model for the predicting the toxicity of the generated molecules and constituting an integral part of the fitness score function. As this task is newly introduced, there are no existing baselines available for comparison. 

% We divided this task into 3 sub tasks to test the abilities of LEOMol in single and multi objective optimization tasks. The sub tasks are 1) Toxicity optimization here the only objective is to generate Non-Toxic molecules 2) QED optimization here the only objective is to generate molecules with higher QED scores 3) Toxicity optimization + QED optimization here the objective is to generate Non-Toxic molecules with higher QED scores. We evaluate this tasks mostly based on 3 metrics \% of Non-Toxic Molecules , Diversity of molecules and standard deviation , mean values of QED scores of the generated molecules. We generated 100 molecules for each of the sub task and fitness score will be different for each one.

\begin{equation}
\textbf{fitness score} = 
\begin{cases}
    % - g_{\theta}(M), & \text{for TO}  \\
    P_{s}(M), & \text{for QED}  \\
    P_{s}(M) - g_{\theta}(M), & \text{for TO + QED}  \\
\end{cases}
\end{equation}

where $g_{\theta}(\cdot)$ is the network  model for predicting toxicity of the generated molecule. where $g_{\theta}(M)$ = 1 indicates $M$ is a Toxic molecule and $g_{\theta}(M)$ = 0 indicates $M$ is a Non-Toxic Molecule. 

Based on the outcomes presented in Table \ref{table:4}, it is evident that, in the absence of the toxicity constraint, the molecules were appropriately optimized to exhibit higher QED scores. However, only 18\% of these optimized molecules demonstrated non-toxic properties, while the remaining 82\% were found to be toxic and unsuitable. Upon the imposition of the toxicity constraint into the fitness function, comparable QED scores were maintained, resulting in the generation of entirely non-toxic molecules with commendable diversity.
Consequently, these results indicate the importance of validating models and approaches based on their ability to generate non-toxic molecules. 

% \textcolor{red}{It is noteworthy that even models adept at optimizing molecules may encounter challenges in producing a substantial percentage of non-toxic molecules, thereby undermining the overall objective.} 
% \textcolor{orange}{We referenced this in the beginning of this section}
% The performance of the LEOMol model across various configurations is presented in Table \ref{table:4}. Notably, in both single objective settings, our approach adeptly achieves its specified objectives. Even in the multi-objective scenario, LEOMol demonstrates the capability to generate 100\% non-toxic molecules, maintaining high diversity and favorable QED scores. This underscores the effectiveness of LEOMol, particularly in practical applications within drug discovery where multi-objective optimization is a prevalent requirement.
% Table \ref{table:4} demonstrates the LEOMol performance in various settings. We can observe in both the single objective settings our approach does a very good job which it was set for. Even in the Multi objective scenario LEOMol is able to generate 100\% Non Toxic Molecules with high diversity and QED scores. This shows the ability of LEOMol to perform effectively even in multi-objective scenario which is the more practical use case in Drug Discovery.
\begin{figure}[!t]
    \begin{subfigure}
    \centering
    \includegraphics[width=0.8\linewidth]{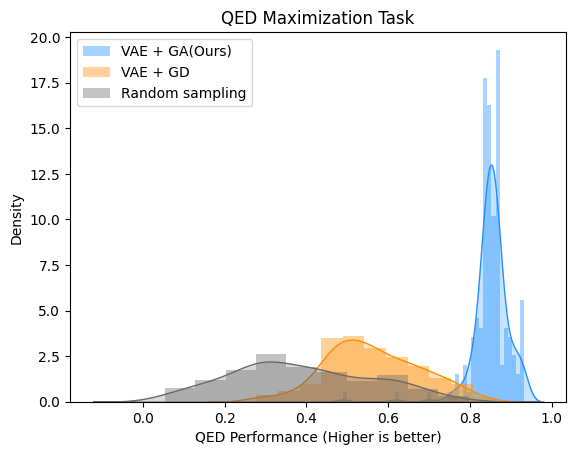}
    \end{subfigure}
    \begin{subfigure}
    \centering
    \includegraphics[width=0.8\linewidth]{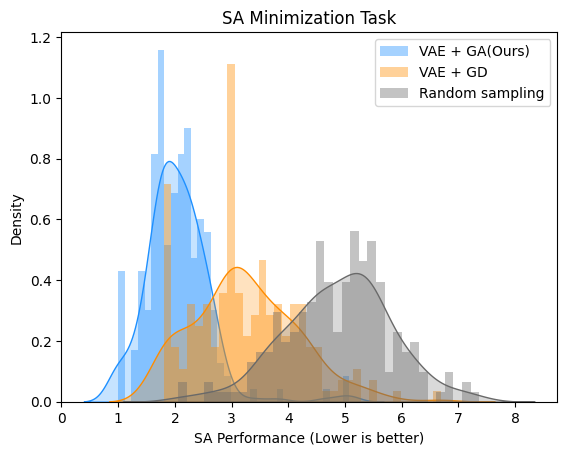}
    \end{subfigure}
    \vspace{-0.2cm}
    \caption{Density plots illustrating property scores of molecules produced through Genetic Algorithm, Gradient Descent Algorithm searches and Random sampling technique within the VAE latent space for tasks involving QED score Maximization and SA score Minimization.}
    % \textcolor{orange}{why did we make this bold? @sid?}
    % The visual representation underscores the evident superiority of our method in generating the desired molecules when contrasted with the Gradient Descent Algorithm (GDA) and random sampling approaches.}}
    \label{Ablation}
\end{figure}

\section{Ablation Study}

To empirically evaluate the effectiveness of LEOMol, a comparative analysis was conducted involving three distinct approaches: a) employing Genetic Algorithm (GA) search within the Variational AutoEncoder (VAE) latent space, b) utilizing Gradient Descent Algorithm (GD) search, and c) employing random sampling within the VAE latent space. Notably, in the GD search approach, the utilization of RDKit is precluded due to its non-differentiable nature. Consequently, a network was trained to predict molecular properties by utilizing the VAE decoder output as input.
The experimental analysis focused on two critical tasks in the drug discovery timeline: maximizing Quantitative Estimation of Drug-likeness (QED) and minimizing Synthetic Accessibility (SA). It is essential to underscore that the same VAE, pretrained on the ZINC250K dataset, was employed for both tasks, ensuring consistency in the experimental setup.

The results are illustrated in Figure \ref{Ablation}. For both tasks, a total of 200 molecules were generated using three distinct approaches. Density plots were then constructed based on the property scores of the molecules. The outcomes demonstrate the superior efficiency of the Genetic Algorithm (GA) search in producing the desired molecules compared to the Gradient Descent Algorithm (GD) search and random sampling technique.
% The findings are depicted in Figure \ref{Ablation}. In conducting both tasks, we generated 200 molecules employing the three distinct approaches. Subsequently, we constructed density plots based on the property scores of the molecules. The results distinctly indicate the superior efficiency of the Genetic Algorithm (GA) search in generating the desired molecules in comparison to the Gradient Descent Algorithm (GD) search and random sampling technique.
% It is crucial to note that the diminished performance of the GDA search is not attributable to its search algorithm per se, but rather to the use of empirical prediction models. These models exhibit limited capacity to generalize the decoder outputs effectively, resulting in inaccurate property predictions that, in turn, contribute to the inefficiency observed in the optimization search.}
% The results are illustrated in the Figure \ref{Ablation}. Here we generated 100 molecules for both the tasks using the two approaches. we plotted the density plots of the molecules according to its property score. We can clearly observe GA search is very efficient in generating the desired molecules compared with the GDA search. Note that the under performance of the GDA search is not due to its search algorithm rather the utilization of the empirical prediction models. These models cannot generalize the decoder outputs much well leading to inaccurate property prediction which leads to the inefficient optimization search.

\section{Discussion and Conclusion}
 We introduced LEOMol, a lightweight generative modeling framework designed for efficient \textit{de novo} drug design. The Evolutionary Algorithm in LEOMol exhibits remarkable flexibility, enabling the seamless integration of non-differentiable property predictors. This capability, which is constrained in previous gradient-based latent search optimization methods, allows for a more comprehensive approach for molecular optimization.
Our in-depth experimental analysis demonstrates the consistent superiority of our approach over prior state-of-the-art models in both property optimization and property targeting tasks. Notably, the findings highlight the higher effectiveness of our Genetic Algorithm (GA) search in generating desired molecules compared to both Gradient Descent Algorithm (GD) search and random sampling techniques. We emphasized the significance of incorporating toxicity as a constraint in the evaluation of optimization methodologies.
% \textcolor{red}{We underscored that evaluating models without considering the toxicity of generated molecules can undermine the overall optimization objective.} 
The application of LEOMol in Drug Discovery shows promise due to its ability to rapidly generate desired molecules while maintaining a 100\% validity score through the use of SELFIES. This capability is expected to accelerate both hit generation and lead optimization stages, effectively and rapidly advancing the forefront of drug discovery. Incorporating advanced techniques for selecting parents from elite indices and exploring alternative Evolutionary Algorithms for latent space exploration, are a promising direction for future work.

\bibliographystyle{plain} 
\bibliography{references} 

\end{document}